\documentclass[preprint,aps]{revtex4}
\usepackage{amssymb}

\usepackage{ gensymb }
\usepackage{graphicx}
\usepackage{dcolumn}
\usepackage{bm}

\begin{document}

\begin{titlepage}

\title{Bending Strain Engineering of Spin Transport in Quantum Spin Hall Systems: Topological Nano-mechanospintronics}

\author{Bing Huang$^{1,2}$, Kyung-Hwan Jin$^{2}$, Bin Cui$^{2}$, Feng Zhai$^{3}$, Jiawei Mei $^{1,2}$, and Feng Liu$^{2,4}$\footnote{Email: fliu@eng.utah.edu}}

\address{$^1$ Beijing Computational Science Research Center, Beijing 100193, China}
\address{$^2$ Department of Materials Science and Engineering, University of Utah, Salt Lake City, UT 84112, USA}
\address{$^3$ Department of Physics, Zhejiang Normal University, Jinhua 321004, China}
\address{$^4$ Collaborative Innovation Center of Quantum Matter, Beijing 100084, China}

\date{\today}

\begin{abstract}
Quantum spin Hall (QSH) system can exhibit exotic spin transport phenomena, mediated by its topological edge states. Here a novel concept of bending strain engineering to tune the spin transport properties of a QSH system is demonstrated by both model and first-principles calculations. Interestingly, we discover that bending strain can be used to mitigate the spin conservation of a QSH system to generate a non-zero spin current (SC), meanwhile the preservation of time reversal symmetry renders its edge states topologically protected to transport robust SC without back scattering. This novel physics mechanism can be applied to effectively tune the SC and spin Hall current in a QSH system by control of its bending curvature. Furthermore, the realization of QSH systems with controllable curvature can be achieved by the concept of ``topological nanomechnical architecture". Taking Bi/Cl/Si(111) as a material example, we demonstrate that the relative spin orientations between two edge states of a Bi/Cl/Si(111) film can indeed be tuned dramatically by its self-bending behaviors induced by the pre-designed inherent strain. Therefore, this concept of ``bending strain engineering of spins" via topological nanomechanical architecture affords a promising route towards the realization of topological nano-mechanospintronics.
\end{abstract}

\maketitle

 \draft

\vspace{2mm}

\end{titlepage}

\section{Introduction}

A long-standing interest in spintronics is generating and transporting spin current (SC) in condensed matter systems. In the past decades, significant process has been made towards realization of highly polarized SC with ferromagnetic materials\cite{Fiederling-1999, Schmidt-2000}, in which SC is strongly coupled with charge current (CC). An ideal ferromagnetic material is half-metal, in which one spin channel is conducting while the other one is insulating\cite{Park-1998, Son-2006}. The discovery of spin Hall current (SHC) that is decoupled from CC\cite{Murakami-2003, Sinova-2004, Kane-2005, Bernevig-2006} has opened up exciting new opportunities for spin transport, because it is expected that the transport of SHC has much smaller energy dissipation compared to that of conventional SC generated by ferromagnetic materials. Quantum spin Hall (QSH) system can exhibit exotic spin transport properties. For a conventional {\it flat} QSH insulator, there are two basic properties, time reversal symmetry (TRS) and spin conservation, which are of special interest. TRS renders the edge states of a QSH insulator topologically protected to transport robust SC without back scattering. However, spin conservation mandates that there is no net SC in a QSH system\cite{Kane-2005}. Although discovering new mechanism to control the SC and/or SHC in a QSH system is of great importance for spintronics, its development is still at its infancy.

Strain engineering has been developed as a well-established approach to enhance the performance of electronic devices, such as Si transistors\cite{Bir-1974}, by tuning band structure and carrier mobility of semiconductors\cite{Lee-2005, Ieong-2004}. Recently, strain engineering has been extended to create novel physical phenomena in 2D materials, e.g., pseudo-magnetic fields\cite{Guinea-2010, Levy-2010} and superconductivity\cite{Si-2013} in graphene. Moreover, strain engineering has also been exploited in materials fabrication through strain induced self-assembly of nanostructures in heteroepitaxial growth of thin films\cite{Yang-2004, Prinz-2000, Schmidt-2001} and most recently through strain partitioned nanomembranes and nanomechanical architecture\cite{Huang-2011}.

In the same spirit of conventional strain engineering of electronic properties, the strain engineering of topological properties has been recognized\cite{Bernevig-2006, Zhu-2012}, because strain changes the bulk band gap of TIs inducing topological phase transitions. Usually, the form of strain considered is tensional strain via lattice expansion/compression. In this article, we explore a new form of ``bending strain engineering" to tune the spin transport of QSH edge states by curvature effect. We demonstrate that for a QSH system under bending strain, curvature preserves its TRS but mitigates spin conservation, so that a spin torque occurs to generate a non-zero SC, which can make this system working as a \emph{topological half-metal} under a bias. This novel idea can further be applied to control the magnitude of SHC in a QSH system by control of its bending curvature, which has not been achieved in a QSH system before. In terms of material design, we suggest a possible approach to grow the self-bending QSH systems via the concept of ``topological nanomechnical architecture", as demonstrated by the material example of Bi/Cl/Si(111) nanofilm, which may open a novel route towards the realization of topological nano-mechanospintronics.

\section{Results}

\begin{figure}[tbp]
\includegraphics[width=10.0cm]{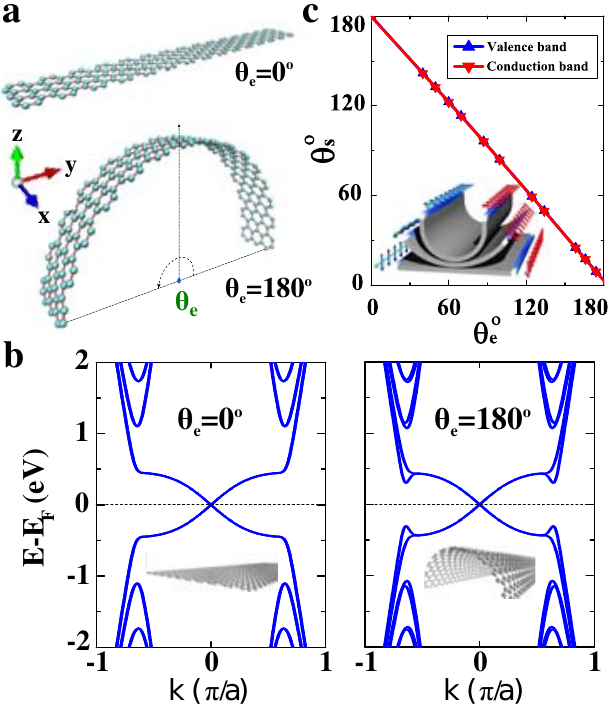}
\caption{(\textbf{a}) Flat and curved structures of a zigzag-edge ribbon with hexagonal lattice. The $\theta_e$, defined as the center angle between the two edges of the curved structure, effectively represents the bending curvature. The coordinate axes are also indicated. (\textbf{b}) Band structures of flat ($\theta_e$=0$^o$) and curved ($\theta_e$=180$^o$) ribbons with 40 atoms/unit cell. (\textbf{c}) Spin angles $\theta_s$ (for both conduction and valence bands) of two double-degenerated edge states (at the same \emph{k} momentum, as marked in \textbf{b}) as a function of $\theta_e$; Inset: schematics of a QSH film and their two edge spin orientations at different bending angles at the same \emph{k} momentum.}
\end{figure}

\textbf{Model of curved QSH systems.}
We start from a curved QSH system on a hexagonal lattice as shown in Fig. 1a. We define a center angle between the left and right edge of a curved ribbon, $\theta_e$, to represent the magnitude of bending curvature. Following Kane and Mele\cite{Kane-2005}, a QSH Hamiltonian generally contains three terms, $H={H_0}+H_R+H_{so}$. The first term $H_0$ defines the band structure. Assuming a sufficiently large spin-orbit coupling (SOC), the gap is insensitive to small changes in hopping or a small staggering potential, then the bending induced change in $H_0$ is negligible. The second Rashba term, associated with the existence of an electric field perpendicular to the plane, is also expected not to change significantly upon bending. Thus, our central attention will be the third term of SOC. Especially, bending changes the directions of orbital angular momenta, which in turn changes the spin directions subject to the spin-momentum locking property. As we will show below, mechanical bending can generate non-zero spin conductance. All the new features of curved QSH systems are intrinsic, which will have profound effects on spin transport properties, independent of specific QSH materials considered.

Concretely, we study a simplified ($p_{x}$, $p_{y}$) four-band model Hamiltonian in a hexagonal lattice\cite{Wu-2008, Liu-2013} (Fig. 1a, upper panel). Neglecting the Rashba effect, $H={H_0}+H_{so}$, we have
\begin{eqnarray}
  {H_0} &=& \mathop \sum \limits_{n,\alpha } {\varepsilon _n}c_{n\alpha }^\dag {c_{n\alpha }} - \mathop \sum \limits_{\langle m,n\rangle ,\alpha } {t_{m,n}}\left( {c_{m\alpha }^\dag {c_{n\alpha }} + h.c.} \right),\\
  {H_{so}}&=&\mathop \sum \limits_{\langle \langle m,n\rangle \rangle } i{t_{so}}\vec{S}\cdot \vec{L} =i{{\lambda _{so}}}\mathop \sum \limits_{\langle \langle m,n\rangle \rangle } \vec{\sigma}\cdot \left({\vec{{e_m}}\times \vec{{e_n}}}\right).
\end{eqnarray}
In $H_{0}$, $c_{n\alpha}^\dag ({c_{n\alpha}})$ creates (annihilates) an electron on atomic orbital $n$ with spin $\alpha$; $\varepsilon_n$  and $t_{m,n}$ are electroinc on-site energy and hopping integral, respectively. In $H_{so}$, $\vec{S}$ and $\vec{L}$ are the spin and orbital angular momentum operators, respectively; $t_{so}$ and $\lambda_{so}$ are constants, defining the SOC strength; $\vec{\sigma}$ is the Pauli vector and $\vec{e_m}$ is the unit vector of $p_x$ or $p_y$ orbital.  In the usual flat system, $\theta_e=0$\degree, $\vec{e_m}\times\vec{e_n}$=$\pm\vec{e_z}$, and spins lie strictly along the $z$ direction. When $\theta_e\neq0$\degree~ upon bending, all the physical observables, e.g., $p_{x,y}$ orbitals, $\vec{e}_m$ and spin directions $\vec{\sigma}$ in Eqs. (1) and (2),  are rotated accordingly by an angle of $\phi$ relative to $x$-axis.

Fig. 1b shows the band structures of flat ($\theta_e$=0$^o$) and curved ($\theta_e$=180$^o$) QSH ribbons, calculated using the TB parameters of $V_{pp\sigma}=$6.38 eV, $V_{pp\pi}=$-2.66 eV \cite{Slater-1954, Tomanek-1988} and $\lambda_{so}$=0.9 eV. The topological Dirac edge states are clearly seen around Fermi level in both cases, which are double-degenerated. For $\theta_e$=0$^o$, the two edge states have opposite spin orientations along $\pm z$ axis, and Pauli matrix of edge states $\vec{\sigma}$ can be described by $\sigma_z$ basis. The spin angle $\theta_s$, defined as the angle between the spin vectors of two edge states, is 180$^{o}$ (Fig. 1c) for $\theta_e$=0$^o$. For $\theta_e$$\neq$0$^o$, bending shows little effect on the shape and degeneracy of edge states, but it significantly changes spin orientation, as shown in Fig. 1c. Upon bending, $\vec{\sigma}$ can be expressed as a linear combination of $\sigma_y$ and $\sigma_z$, and the net spin direction in the whole system is along $y$ axis which can be expressed in $\sigma_y$ basis. The larger the $\theta_e$, the larger the net $\sigma_y$ spin component. Our calculations establish a simple relationship between $\theta_s$ and $\theta_e$ as $\theta_s$+$\theta_e$=180$^o$.

\textbf{Spin transport in curved QSH systems.} Curvature does not remove TRS in curved QSH systems, since spin/charge currents with opposite polarity still propagates in opposite directions along the edges, as shown in Fig. 2a ( from left to right: $\theta_e$ increases from 0$^o$ to 180$^o$) and also reflected by the unchanged edge band structures (Fig. 1b). However, curvature mitigates spin conservation; spins are no longer conserved along the edges, e.g., they adiabatically rotate on the curved sides of edges (Fig. 2a), which is expected to  modifies non-equilibrium spin transport properties in curved QSH systems under a bias. Specifically, edge spin rotation in Fig. 2a is achieved by creating an $S_y$ component in addition to $S_z$. At two opposite edges of a QSH ribbon, the $S_z$ components are antiparallel but the $S_y$ components are parallel to each other along the same charge current flow direction. Consequently, a conventional (flat) QSH insulator conducts only CC but not SC because only the $S_z$ component is present, while a curved QSH insulator can conduct both CC and SC arising from the emergence of $S_y$ component.

\begin{figure}[tbp]
\includegraphics[width=12.0cm]{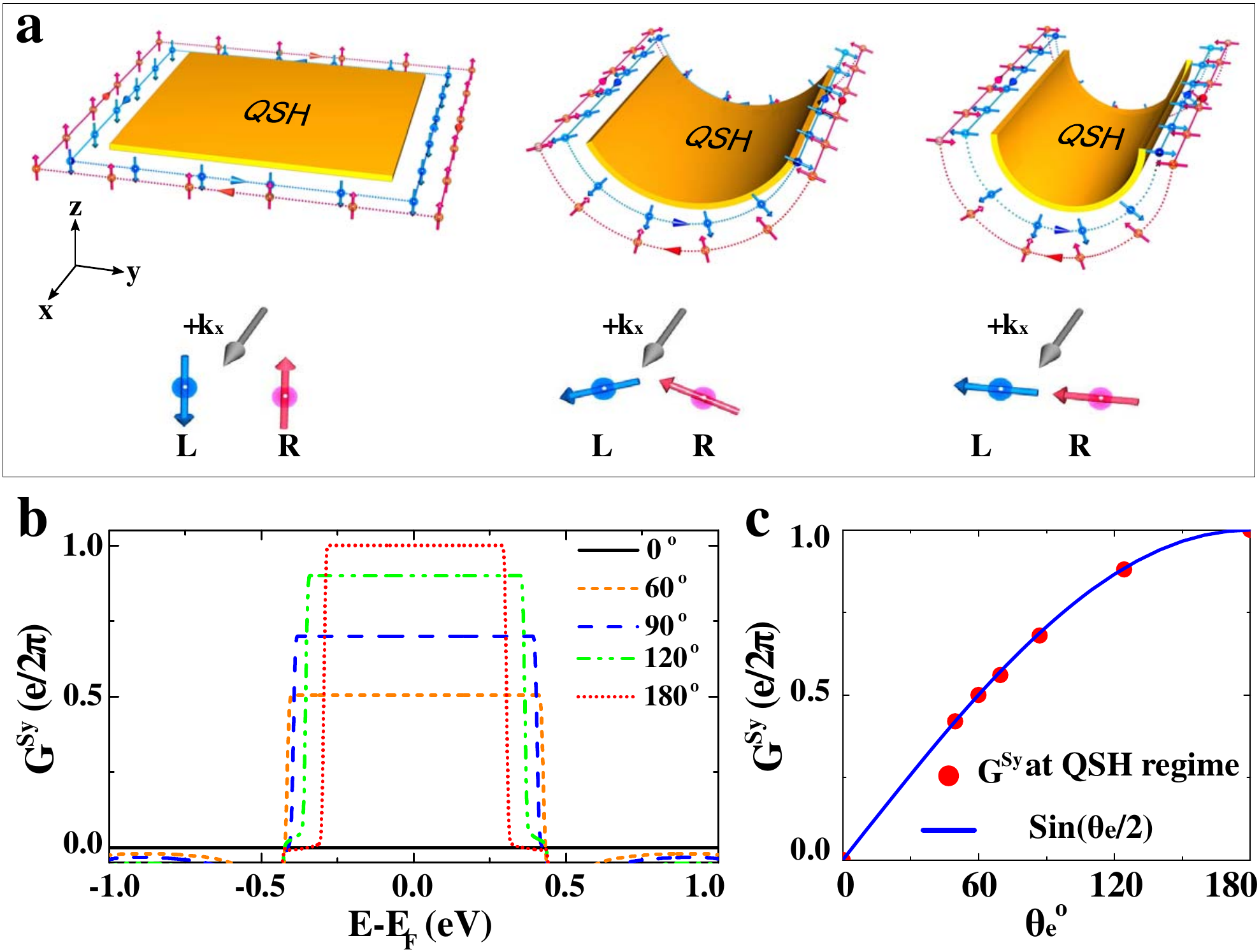}
\caption{(\textbf{a}) Schematic diagrams of spin current and charge current flowing along the edges as the $\theta_e$ increases from 0$^o$ (left) to 180$^o$ (right). A pair of edge states counter propagate along all four edges subject to TRS. The spins rotate adiabatically along the curved edges. The highlight of spin directions at the two opposite edges under the same charge current flow direction +k$_x$ is shown in the bottom. (\textbf{b}) Calculated spin conductance $G^{S_y}$ ($G^{S_x}$=$G^{S_z}$=0) for the QSH ribbons with different $\theta_e$ in a two-terminal device setting. (\textbf{c}) The values of $G^{S_y}$ in the QSH regime (plateau region) in (\textbf{b}) as a function of $\theta_e$, which can be perfectly described by the equation of $G^{S_y}=\sin(\theta_e/2)\cdot\frac{e}{2\pi}$.}
\end{figure}

Quantitatively, we can calculate the charge and spin transmission coefficients of QSH ribbons with different $\theta_e$ in a two-terminal device setting using non-equilibrium Green's function formalism in the linear-response regime\cite{Datta-1995} as
\begin{equation}
T\left( E \right) = {\rm{T}}r\left[ {{{\widehat\sigma}_\alpha }{{\rm{\Gamma}}_L}{\mathbb{G}^r}\left(E\right){{\rm{\Gamma}}_R}{\mathbb{G}^a}\left(E \right)}\right],
\end{equation}
where ${{\rm{\Gamma}}_{L/R}} = i\left[ {{{\rm{\Sigma }}_{L/R}} - {\rm{\Sigma }}_{L/R}^ + } \right]$
indicates the interaction between a central scattering area and left/right lead, whose self-energy is ${{\rm{\Sigma}}_{L/R}}$. ${\mathbb{G}^{r/a}}\left(E\right)$ is the retarded/advanced Green¡¯s function; ${\widehat\sigma _\alpha}$, a unit matrix for calculating the charge conductance, is one of the Pauli matrices for spin conductance\cite{Chang-2014}. Both source/drain and central scattering area are made of the same material to reveal the intrinsic transport properties.

At the QSH regime (plateau region), our calculations show that the charge conductance $G$ is insensitive to curvature and remains at its quantized value, as shown in Fig. S1 \cite{SM}, but the spin conductance $G^{S_y}$ (here $G^{S_x}$=$G^{S_z}$=0) becomes $\theta_e$-dependent and no longer quantized. As shown in Fig. 2b, $G^{S_y}$ gradually increases from 0 ($\theta_e$=0$^o$) to $e/2\pi$ ($\theta_e$=180$^o$) with the increasing $\theta_e$ in the QSH regime, consistent with the increased $S_y$ component (Fig. 1c). The spin conductance $G^{S}$ can only take the forward-propagating edge modes, whose direction is given by the direction of external bias.

Furthermore, we can make some general arguments to illustrate the effects of $\theta_e$ on the $G^{S}$ of a QSH system.  Actually it is legal for us to implement a local coordinate (unitary) transformation on the bending curved QSH system
\begin{eqnarray}
  \label{eq:lct}
   H_C\rightarrow \tilde{H}=  \mathcal{R}^\dag H_C\mathcal{R},\quad\mathcal{R}=\prod_mR_m^dR_m^s,
\end{eqnarray}
such that we can transform our curved system $H_C$ into a \emph{flat} one $\tilde{H}$. Here $R_m^d$ is spin-independent deformations and $R_m^s=e^{i(S_x)_m\phi(m)/\hbar}$ is the spin rotation.  If we ignore the Rashba effect $H_R$, it is easy to see that the total rotated spin z-component $\tilde{S}_z=\mathcal{R}^\dag S_z\mathcal{R}$ is conserved, $[\tilde{S}_z, \tilde{H}_C]=0$, and the spin Chern number $\tilde{C}_s$ in the flat system  is well defined\cite{Sheng-2006}. The edge spin conductances at left and right  have only non-zero $z$-component $g^{\tilde{S}_z}_{L,R}=\tilde{C}_s/2\cdot\frac{e}{2\pi}$. These results are well established for a flat QSH system. If we go back to the original curved reference frame, we have the edge spin conductance, $g^{S_y}_{L,R}=\sin(\phi_{L,R}/2)\tilde{C}_s/2\cdot\frac{e}{2\pi},\quad g^{S_z}_{L,R}=\cos(\phi_{L,R}/2)\tilde{C}_s/2\cdot\frac{e}{2\pi}$. By definition, the net spin conductance is given as $ G^{S_y}=(g^{S_y}_{L}-g^{S_y}_{R})/2\cdot\frac{e}{2\pi},\quad G^{S_z}=(g^{S_z}_{L}-g^{S_z}_{R})/2\cdot\frac{e}{2\pi}$.
Therefore, in our curved system, $\phi(L)=-\phi(R)=\theta_e, \tilde{C}_s=2$, then spin conductance is
\begin{eqnarray}
  G^{S_y}=\sin(\theta_e/2)\cdot\frac{e}{2\pi},\quad G^{S_z}=0.
\end{eqnarray}
This conclusion agrees well with our calculations, as shown in Fig. 2c. We want to emphasize that all these new features of curved QSH systems are intrinsic, having profound effects on spin transport properties independent of specific QSH materials considered.

\textbf{Tunable SC and SHC in curved QSH systems.} Based on the same physical mechanism, curvature can also modify the SHC in curved QSH systems, because the SHC is only contributed by the $S_z$ components which decreases with the increased $S_y$ components, and the quantization of the spin Hall conductance in a QSH system is only guaranteed when $S_z$ is conserved\cite{Kane-2005-2}.

More generally, we provide a comparison between the charge and spin transport properties of curved QSH devices and those of conventional (flat) QSH devices in both two- and four-terminal settings within the Landauer-B\"{u}ttiker\cite{Landauer} framework, as shown in Fig. 3. In terms of transport, with two terminals, the curved QSH (Figs. 3b and 3c, upper panel) device conducts both CC and SC [0 to (e/2$\pi$)V], which is significantly different from the flat QSH device that conducts only CC (Fig. 3a, upper panel). A curved QSH device can effectively work as a \emph{topological half-metal} for spin injection, i.e., it transports topologically protected completely spin-polarized charge current, and the density of SC can be tuned by the curvature. With four terminals, the flat QSH device conducts a longitudinal CC ($I_l$) and a transverse SHC ($I^s_t$) (Fig. 3a, lower panel), while the curved QSH device with $0<\theta_e<180^o$ (Fig. 3b, lower panel) conducts both longitudinal CC $I_l$ and SC $I^s_l$ (contributed by $S_y$ component) as well as a transverse SHC $I^s_t$ (contributed by $S_z$ component). Interestingly, $I^s_t$ ($I^s_l$) continues to decrease (increase) with increasing $\theta_e$ subject to the conservation of total spin, $\vec{S}=\vec{S_y}+\vec{S_z}$, and finally $I^s_t$ vanishes at $\theta_e=180^o$ (Fig. 3c, lower panel).

In terms of robustness against back scattering, the curved and flat QSH devices are the same, as they are protected by TRS. In terms of conductance quantization, charge conductance is integer-quantized in unit of $e^2/h$ in both flat and curved QSH devices, as shown in Fig. S1\cite{SM}. However, spin conductance is only integer-quantized in unit of $e/4\pi$ in the flat but not in the curved QSH device, hence in the latter the spin conductance, arising from the $S_y$ components, is not conserved for different $\theta_e$ (Fig. 2c). Therefore, curvature, induced by bending strain, can be employed to dramatically tune the topological SC and SHC in the a curved QSH system for various spintronics applications.

\begin{figure}[tbp]
\includegraphics[width=14.0cm]{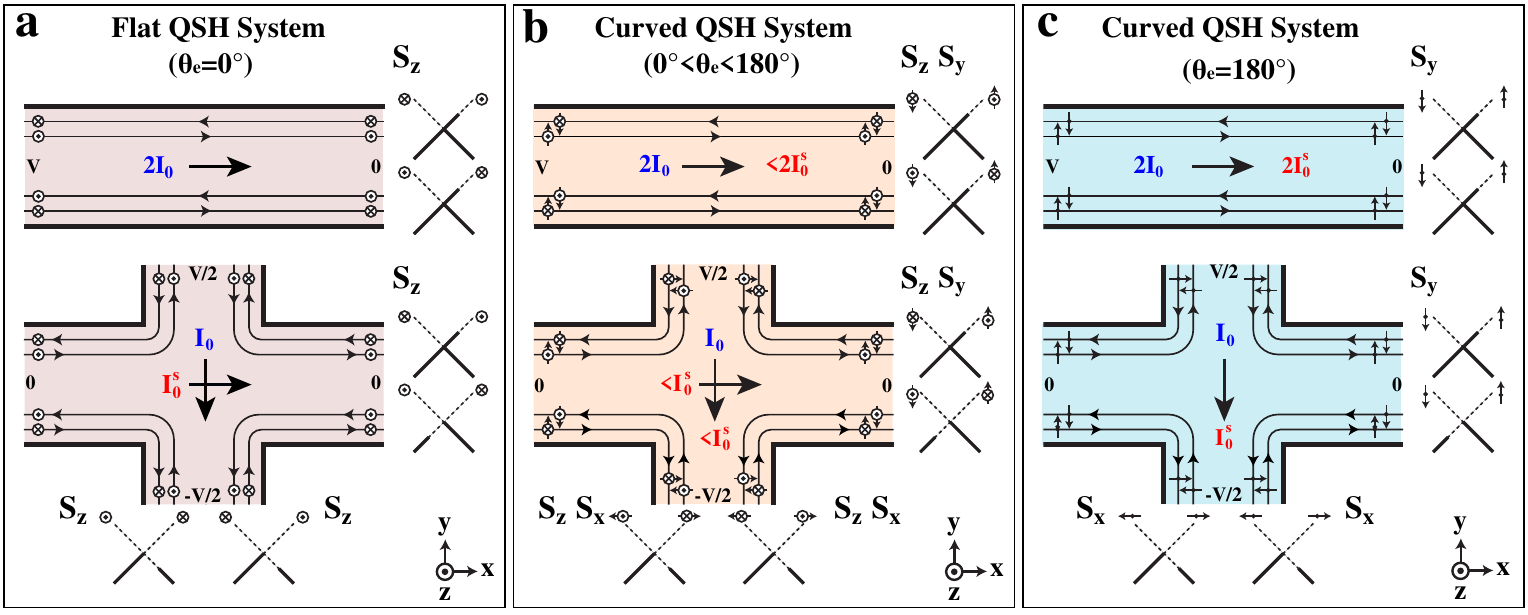}
\caption{Comparison of two-terminal and four-terminal measurement geometries for a (\textbf{a}) flat QSH system ($\theta_e$=0$^o$), (\textbf{b}) curved QSH system ($0<\theta_e<180^o$), and (\textbf{c}) curved QSH system ($\theta_e=180^o$). The arrows indicate the charge current $I$ and spin current $I^s$ and their flow directions. The unit of $I$ and $I^s$ are $I_0=(e^2/h)V$ and $I_0^s=(e/4\pi)V$, respectively. The diagrams to the right and bottom indicate population of the edge states.}
\end{figure}

\textbf{Concept of topological nanomechanical architecture.} A practical idea to realize bent QSH systems is nanomechanical architecture of strained nanofilms, which has been proven a powerful method to fabricate nanomembranes, nanotubes, partial or half nanotubes, and nanocoils\cite{Prinz-2000, Schmidt-2001, Huang-2005}. The general process of nanomechanical architecture proceeds with growth of strained nanofilms on a sacrificial substrate followed by patterning of release (through removal of the sacrificial substrate) of the nanofilms, which will roll-up into different tubular shapes as pre-designed by strain engineering (see our classical MD simulation in Video V1\cite{SM} for this concept). Suppose one can apply the same process to a QSH nanofilm, then a new type of strain engineering of topological boundary states is realized to tune the edge spin orientations in a controllable manner.

Furthermore, it is a parallel process that can facilitate mass production of identical partial cylindrical QSH arrays\cite{Chun-2008}, which will function ideally as a robust spin injector device with high spin current density, as demonstrated in Fig. S2, while spin polarization can be switched by changing the bias direction. Compared to the traditional magnetic materials, e.g., ferromagnetic metals, the QSH system based spin injectors are topologically protected, robust against structural distortion or impurity scattering; the helical Dirac edge states support also ultra-fast SC transport.

\textbf{Curvature effect in Bi/Cl/Si(111) nanofilms.} To demonstrate the feasibility of the above concept, we have further performed first-principles calculations to study the evolution of topological edges states of a QSH Bi/Cl/Si(111) nanofilm under self-bending driven by the nanomechanical architecture process. It has been predicted that a surface based QSH state forms in a hexagonal Bi overlayer deposited in the halogenated Si(111) surface, i.e., Bi/Cl/Si(111)\cite{Zhou-2014}. If one grows a ultrathin Si(111) film on a sacrificial SiO$_2$ substrate before Cl adsorption and Bi deposition, then the resulting Bi/Cl/Si(111) nanofilm is readily subject to the nanomechanical architectural process, sell-rolling into a tubular shape (including a partial cylinder) upon releasing from the underlying SiO$_2$ substrate.

The Si(111) surface functionalized with one-third monolayer (ML) of Cl exhibits a $\sqrt{3}\times\sqrt{3}$ reconstruction\cite{Buriak-2012, Dev-1985}. When 1 ML Bi is deposited on the Cl/Si(111) surface, the most stable structure of Bi atoms adopts a hexagonal Bi lattice (Fig. 4a)\cite{Zhou-2014}. The Bi lattice has an in-plane lattice constant of 3.87 $\AA$, $\sim$20\% larger than that of free-standing Bi layer. This gives rise to a large tensile surface stress of 0.12 eV/$\AA$ in the top surface of Bi/Cl/Si(111), obtained from first-principles calculations. On the other hand, the bottom surface of Bi/Cl/Si(111), which might be bare (or H-passivated) during the release process from the underneath substrate, has a much smaller surface stress of -0.04 (or $\sim$0.00 eV/$\AA$). Therefore, there can exist a stress imbalance between the top and bottom surface of Bi/Cl/Si(111) nanofilm which provides a driving force for self-bending. Using the surface stress difference $\Delta\sigma$ as input, we can estimate the bending curvature $\kappa$ of Bi/Cl/Si(111) nanofilm as a function of film thickness \emph{t} using Stoney formula\cite{Stoney-1909} $\kappa  = (6\Delta \sigma )/(C{t^2})$, as shown in Fig. 4b, where $C = E/(1 - {\nu ^2})$ is a constant related to Young's modulus $E$ and Poisson ratio $\nu$ of Si.

\begin{figure}[tbp]
\includegraphics[width=12.0cm]{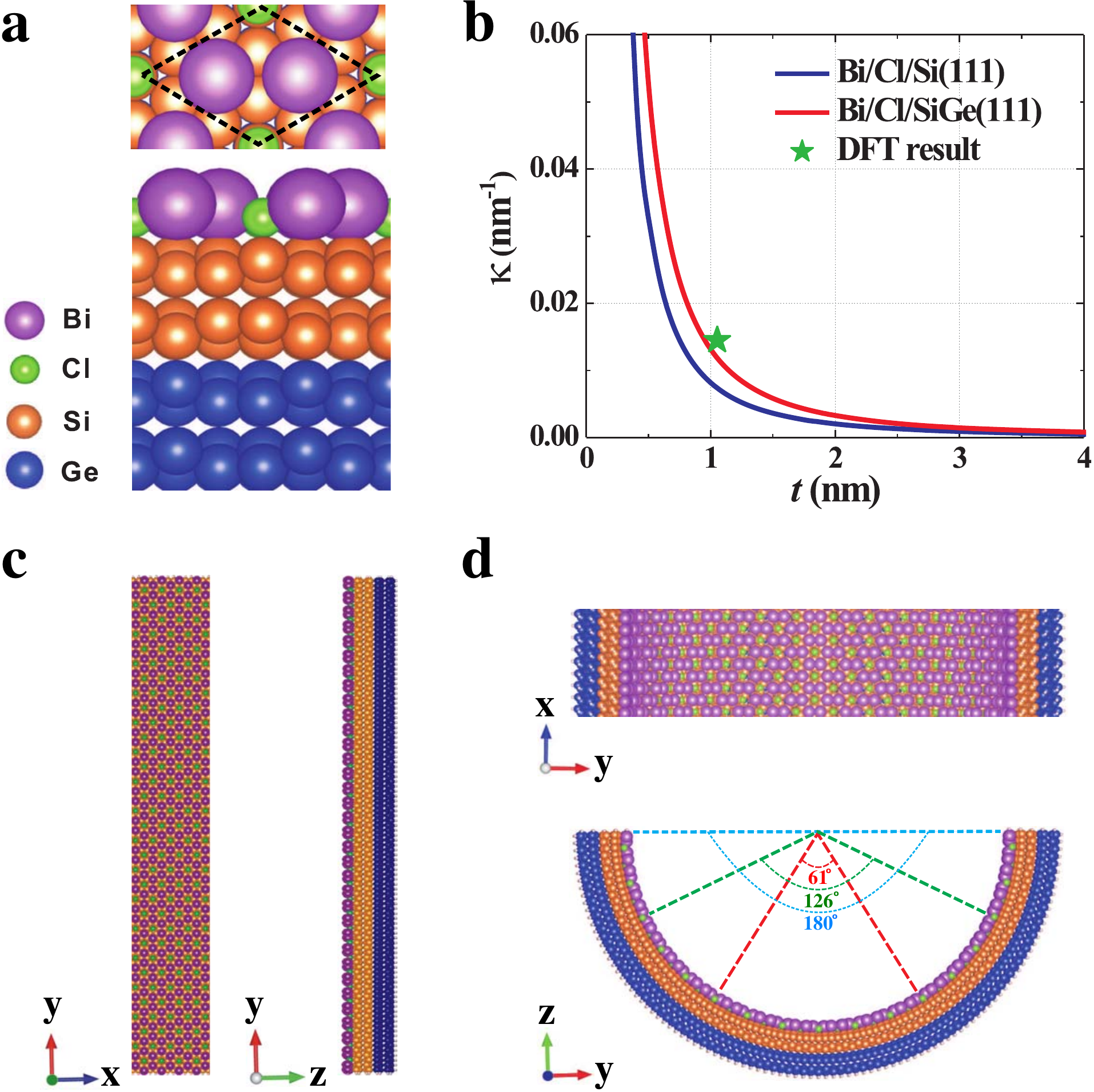}
\caption{(\textbf{a}) Top and side views of a Bi/Cl/SiGe(111) surface. The dashed black lines mark the unit cell. (\textbf{b}) The calculated self-bending curvatures of Bi/Cl/Si(111) and Bi/Cl/SiGe(111) nanofilm as a function of film thickness based on Stoney and Timoshenko formulas, respectively. The DFT simulated bending curvature of a Bi/Cl/SiGe(111) of a 1.05 nm thickness is also shown (a star). (\textbf{c}) Top and side views of a flat Bi/Cl/SiGe(111) surface with 2 atomic layers Si and Ge each. The ribbon edge termination is along the zigzag-edge direction of Bi lattice (edges are passivated by H atoms). (\textbf{d}) Self-bent structure shown in (\textbf{c}) obtained from first-principles total energy relaxation. Three different angles in (\textbf{d}) represent three ribbon widths (along $y$ direction) with different edge bending angles $\theta_e$.}
\end{figure}

As a novel strategy in nanomechanical architecture, besides changing film thickness, another effective way to control the bending curvature of the rolled-up tubular structure is to grow lattice-mismatched multilayer film to partition the amount of misfit strain and tune the driving force for bending. Specifically, SiGe film is often used for this purpose, since Ge lattice is $\sim$4.2\% larger than Si lattice and the growth SiGe film is a well-established technique. To verify this idea, we have taken the bilayer system of Bi/Cl/SiGe(111) film (two atomic layers of Si and Ge each) as an example, and the calculated total imbalanced ``surface" stress in this system is 0.21 eV/\AA, about two times larger than that of Bi/Cl/Si(111). We can estimate the bending curvature of Bi/Cl/SiGe(111) nanofilm as a function of total film thickness \emph{t} from Timoshenko formula\cite{Timoshenko-1925} $\kappa  = (6\Delta \sigma/{E_s}{t^2})\gamma$ and $\gamma  = \frac{{{{(1 + \beta )}^3}}}{{1 + 4\alpha \beta  + 6\alpha {\beta ^2} + 4\alpha {\beta ^3} + {\alpha ^2}{\beta ^4}}}$, where $\alpha=E_f/E_s$, $E_f$ and $E_s$ are Young's modulus of Si and Ge, respectively, $\beta=t_f/t_s$ is the ratio of Si thickness $t_f$ and Ge thickness $t_s$, and $t=t_f+t_s$. The result is shown in Figs. 4b, confirming a larger bending curvature than Bi/Cl/Si (111) system.

Next, we use first-principles calculations to directly simulate the self-bending curvature of a nanoribbon of finite width made of Bi/Cl/SiGe(111) nanofilm, as shown in Figs. 4c-d. For comparison, we again choose two atomic layers of Si and Ge. The edges are along the zigzag edge direction of Bi lattice and passivated with H atoms to remove dangling bonds. The calculated self-bending curvature of Bi/Cl/SiGe(111) is 0.0136 nm$^{-1}$, which agrees quite well with the estimation from Timoshenko formula, i.e., 0.0119 nm$^{-1}$.

\begin{figure}[tbp]
\includegraphics[width=12.0cm]{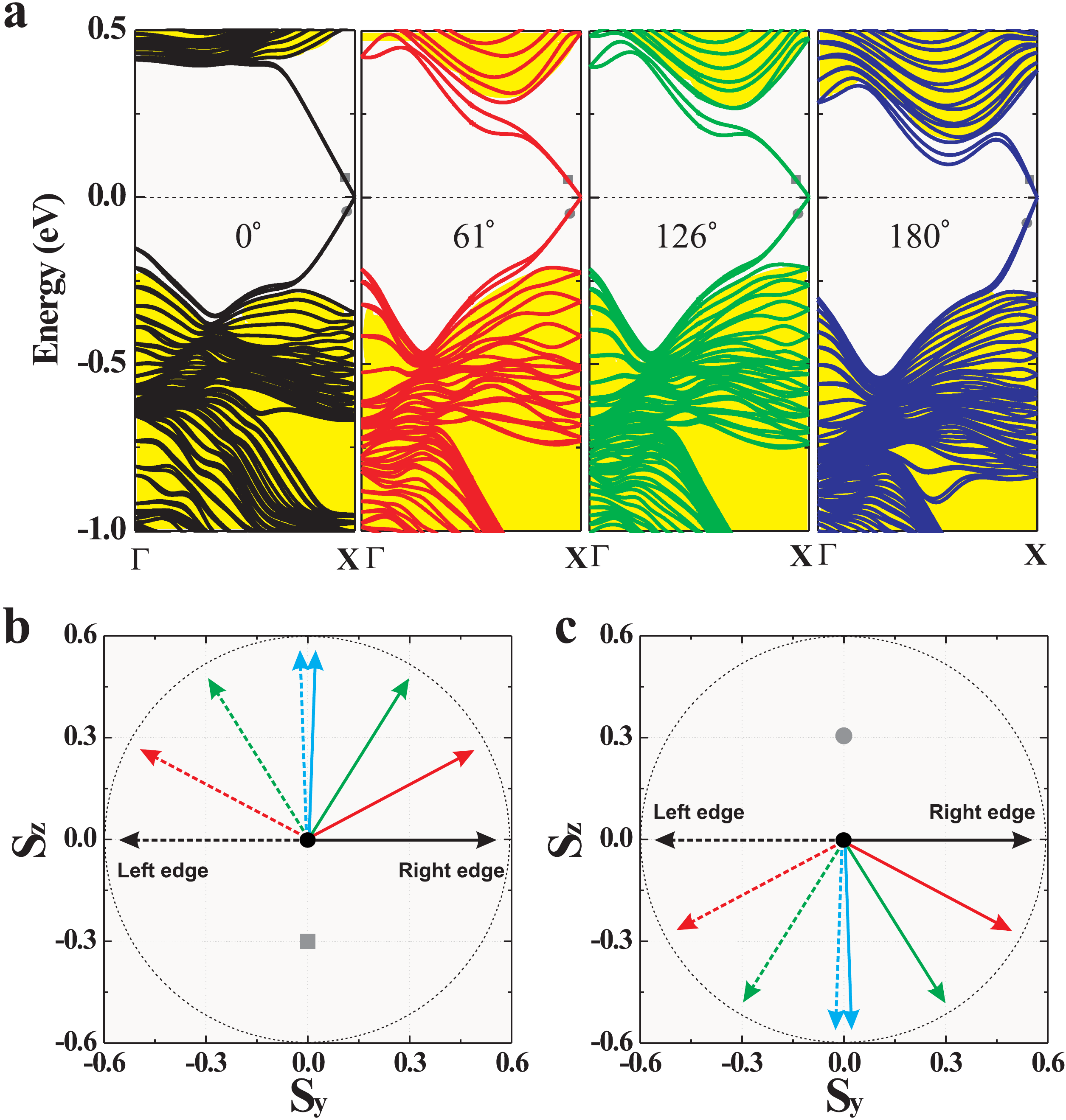}
\caption{(\textbf{a}) First-principles calculated band structures of four Bi/Cl/SiGe(111) ribbons at different bending angle $\theta_e$, 0$^{o}$, 61$^{o}$, 126$^{o}$, and 180$^{o}$, respectively. The bulk bands are marked in yellow region. (\textbf{b}) Spin rotations of conduction band edge states at a \emph{k} momentum slightly off Dirac point (X), as marked in (\textbf{a}). (\textbf{c}) Same as (\textbf{b}) but for valence band edge states.}
\end{figure}

\textbf{Electronic properties of curved Bi/Cl/SiGe(111) nanofilms.} After the self-bending curvature of Bi/Cl/SiGe(111) is determined, three different ribbon widths are used to simulate three different $\theta_e$, which are 61$^o$, 126$^o$, and 180$^o$, as indicated respectively in Fig. 4d, for topological edge state calculations. As Bi $p_z$ orbitals are passivated by the top Si atoms on the substrate, the remaining Bi $p_x+p_y$ orbitals realize a QSH phase, which can be described by a four-band TB model Hamiltonian of Equ.(1) and (2). The calculated band structures for these three $\theta_e$ cases, along with the case of $\theta_e$=0$^o$, is shown in Fig. 5a, where the existence of linearly dispersive Dirac bands crossing the Fermi level indicates a nontrivial band topology. The Dirac edge states persist with bending, as expected from their topological origin to be robust against structural deformation. After bending, the degeneracy of the two edge states are slightly lifted when the energy moves away from the Fermi level because of the broken symmetry.

Figs. 5b-c show the evolution of the spin direction of the conduction and valence edge states slightly off the Dirac point. For $\theta_e$=0$^o$, the spins are aligned normal to the ribbon plane ($z$-axis) and the two edge spins are orientated antiparallel ($\theta_s$=180$^o$) with each other. Upon bending, as shown in Fig. 5b, for conduction band edge states the spins rotate counterclockwise (clockwise) for the left (right) states as $\theta_e$ increases. When $\theta_e$ reaches $\sim$180$^o$, spins are rotated into almost parallel along $y$-axis at both edges ($\theta_s$=4$^o$). Similar behavior is found for valence band edge states, as shown in Fig. 5c. Thus, our first-principles calculations of a real QSH material not only can confirm the concept proposed, but also suggest a promising way to realize novel topological spintronics materials by nanomechanical architecture.

\section{Conclusion}

We have theoretically proposed a novel concept of bending strain engineering of spin transport in QSH systems, which is generally applicable to all QSH materials and especially suited for surface or interface-based QSH states on or inside a thinfilm. It affords a promising route towards realization of robust QSH-based spin injectors with 100\% spin polarization. A curved QSH system may be potentially realized by subjecting a QSH nanofilm to nanomechanical architecture process. Our finding opens a new avenue to topological nano-mechanospintronics, enabling generation and transport of spin current by mechanical bending of a QSH system. It significantly advances our fundamental knowledge of spin transport properties, as well as broadens the scope of nanotechnology into topological materials and devices and vice versa.

\section{Methods}

First-principles calculations based on the density functional theory were performed within the generalized gradient approximation of PBE form for the exchange-correlation of electrons as implemented in the VASP Package\cite{VASP}. The projected-augmented-wave method was used to describe the atomic potentials. The SOC was included at the second variational step using the scalar-relativistic eigen-functions as a basis. A cutoff energy of 450 eV was used for the expansion of wave functions and potentials in the plane-wave basis. Sufficient k-point meshes were used for sampling the Brillouin zone. The atomic structures of all the calculated systems were fully relaxed until the Helmann-Feynman forces were less than 0.02 eV/\AA. To simulate the nanoribbon structures in the plane-wave basis, we employed the supercell method. Both the edge-to-edge and layer-to-layer distances between adjacent ribbons are set $>$20\AA, to eliminate artificial interactions between neighboring cells.

\section{Acknowledgements}

This work was supported by US-DOE (Grant No. DE-FG02-04ER46148) (B.H., K.W.J, F.L.). B.H. also acknowledges support from Chinese Youth one-thousand Talents Program and NSFC (Grant No. 11574024). Computations were performed at DOE-NERSC and CHPC of University of Utah.

\section{Author contributions}

F. L. and B. H. directed the project. B. H., K. J., B. C., F. Z. and J. M. calculated and analyzed the results. B. H. and F. L. wrote the manuscript. All authors discussed the results and commented on the manuscript.

\end{document}